\documentclass[reprint,prl,superscriptaddress]{revtex4-1}
\usepackage[utf8x]{inputenc}
\usepackage{ucs}
\usepackage{amsmath}
\usepackage{amsfonts}
\usepackage{amssymb}
\usepackage{makeidx}
\usepackage{graphicx}
\usepackage{xcolor}
\usepackage{listings}

\begin{document}

\author{Tina Hecksher} \affiliation{Glass \& Time, IMFUFA, Department
  of Sciences, Roskilde University, Postbox 260, DK-4000 Roskilde,
  Denmark}

\title{Linking the dielectric Debye process in 2-ethyl-1-hexanol to
  its density fluctuations.}

\date{\today}

\begin{abstract}
  We provide the first evidence that the puzzling dielectric Debye
  process observed in mono-alcohols is coupled to density
  fluctuations. The results open up for an explanation of the Debye
  process within the framework of conventional liquid-state
  theory. The spectral shape of dynamical bulk modulus of
  2-ethyl-1-hexanol is nearly identical to that of the shear modulus,
  and thus the supramolecular structures believed to be responsible
  for the slow dielectric Debye process are manifested in the bulk
  modulus \emph{in the same way} as in the shear modulus.
\end{abstract}

\maketitle

All liquids can form a glass \cite{Kauzmann1948, Scherer1986a,
  Ediger1996, Angell2000, Debenedetti2001, Dyre2006a, Ediger2012,
  Berthier2016}. The most common route to the glassy state is through
the supercooled liquid state where the liquid gradually becomes more
and more viscous upon cooling and eventually falls out of equilibrium
and forms a glass.  Many viscous liquids display strikingly similar
dynamical behavior regardless of their specific chemistry and this
universality intrigues many physicists \cite{Kauzmann1948,
  Scherer1986a, Ediger1996, Angell2000, Ediger2012, Berthier2016,
  Debenedetti2001, Dyre2006a}. But although the field is old, the
hallmark features of viscous liquids dynamics -- non-Arrhenius
temperature dependence of the viscosity and non-exponential relaxation
-- remain some of the major unsolved puzzles in condensed matter
physics.

A class of liquids that do not conform to the general picture is the
mono-alcohols. Mono-alcohols have become a hot topic in recent years
(see the review by B\"{o}hmer \textit{et al}
\cite{Bohmer2014}). Supercooled mono-alcohols differ from other
viscous liquids in that the dominant process in the dielectric
spectrum is: 1) Close to single-exponential rather following the
stretched exponential form found in most viscous liquids
\cite{Davidson1951, Cole1952, Dannhauser1968}; mono-alcohols thus
follow the prediction of Debye's theory \cite{Debye1929}, and the
process is therefore often referred to as the ``Debye-process''. 2)
Not associated with the structural glass transition \cite{Kudlik1997,
  Hansen1997, Wang2004, Huth2007}, as is usually seen in viscous
liquids where the kinetically defined glass transition temperature
$\tau_\alpha(T_g) = 100$~s correlates well with the calorimetrically
defined $T_g$. The faster process emerging on the high-frequency flank
of the Debye process correlates much better with the structural glass
transition and this process has accordingly been identified as the
alpha process \cite{Kudlik1997, Hansen1997, Jakobsen2008}. 3) Very
intense. The Debye process is often so intense that it must originate
from some structural correlation of several dipoles in the
liquid. Dilution studies \cite{Schwerdtfeger2001,Preuss2012,
  Bauer2013, Gao2013} and studies of structural isomers
\cite{Murthy1996, Hecksher2014} show that these characteristics may be
less distinct, when the Debye process is not well-separated from the
alpha process.

The notion that this intense dielectric signal originates from linear
hydrogen-bonded structures appeared quite early \cite{Dannhauser1968,
  Johari1968, Floriano1989, Iwahashi1993}, and observations of a
pre-peak in the static structure factor \cite{Morineau1998} supports
this idea by identifying structures on a length scale exceeding the
molecular. The different intensities of the Debye process in different
mono-alcohols can be rationalized in terms of these structures being
primarily chain-like -- leading to a large end-to-end dipole moment --
or ring-like -- resulting in a (partial) cancellation of the
individual dipole moments \cite{Dannhauser1968}.


A number of mechanisms have been suggested to account for the slow
dynamics of the Debye process: breaking of hydrogenbonds in a chain
and formation of new chains leading to end-to-end fluctuations of the
dipole moment \cite{Sagal1962, Kalinovskaya2000, Kaatze2002}, dipole
inversion by rotation of the OH-groups \cite{Minami1980}, and the
transient chain model advocated by Gainaru \cite{Gainaru}, where
molecules break off from or add to the ends of the chain which
promotes a slow rotation of the effective dipole moment of the
chain. A fundamental and predictive model for the elusive Debye
process is however still lacking. In particular, these models have
nothing to say about whether or not this intense dielectric should be
manifested in other types of responses. For years the consensus was
that the Debye process is not linked to the mechanical or calorimetric
responses, as summarized in 1997 by Hansen \textit{et al}, the Debye
process ``possesses no counterpart signals in the quantities directly
related to structural relaxation like viscosity and density
fluctuations'' \cite{Hansen1997}. Earlier shear modulus measurements
seemed to confirm this picture \cite{Jakobsen2008}. However, there
were some indications of a mechanical signature of the Debye process
in ultrasonic measurements \cite{Behrends2001, Kaatze2002}, and it was
recently established that the Debye process indeed has a weak shear
mechanical counterpart \cite{Gainaru2014}. The rheological response is
similar to what is observed for short-chain polymers
\cite{Gainaru2014, Gainaru2014a, Hecksher2014, Adrjanowicz2015}. The
detection of a shear mechanical signature of the dielectric Debye
process is difficult because the signal is small -- especially
compared to how dominant the dielectric signal is -- and thus requires
good resolution and accurate measurements.

A natural next step is to look for signatures in thermodynamic
response functions. Could the dielectric Debye process in fact be
coupled to the density fluctuations?  This would open up a new route
for modeling the Debye process. Density and density fluctuations are
the central concepts in liquid state theory, where they define
standard hydrodynamics \cite{Hansen2013}, mode-coupling theory
\cite{Bosse1978, Gotze1992, Das2004}, and general density functional
theory \cite{Evans1979,Rosenfeld1989, Evans1992}.
Moreover, density fluctuations provide a link to classical
thermodynamics, where the state variables are scalar quantities such
as pressure, volume, temperature and entropy. Both shear mechanics and
dielectrics are non-scalar observables, and since a signature of the
Debye process is thus far completely absent in the standard
calorimetric scans, there is no established connection to
thermodynamics.

Experimentally density fluctuations may be probed by measuring the
volume response to a (linear) pressure perturbation. Recently, Dzida
and Kaatze \cite{Dzida2015} published a study comparing the static
adiabatic compressibility and the dielectric relaxation time of a
range of mono-alcohols and showed how the concentration of
hydrogenbonds affects the Debye relaxation time and the static
compressibility differently at room temperature. But from this study
no clear conclusions can be drawn about the dielectric Debye process'
manifestation in the compressibility spectrum.

Here we present complex adiabatic bulk modulus data of the
mono-alcohol 2-ethyl-1-hexanol (2E1H) measured over roughly four
decades in frequency over a range of temperatures close to $T_g$. The
adiabatic bulk modulus is defined as $K_S = V \left( \partial
  P/\partial V \right)_S$, i.e., the inverse of the
compressibility. In addition, the complex shear modulus ($G$) was
measured at the same temperatures and frequencies (spectra are shown
in the Supplementary Material). Both bulk and shear modulus
measurements were carried out in the same closed-cycle custom-built
cryostat with the same measuring electronics \cite{Igarashi2008a,
  Igarashi2008b}, ensuring identical experimental conditions. More
details on the methods can be found in Refs.\
\onlinecite{Christensen1994b} and \onlinecite{Christensen1995}, and in
the Supplementary Material.

\begin{figure}
\includegraphics[width=7.8cm]{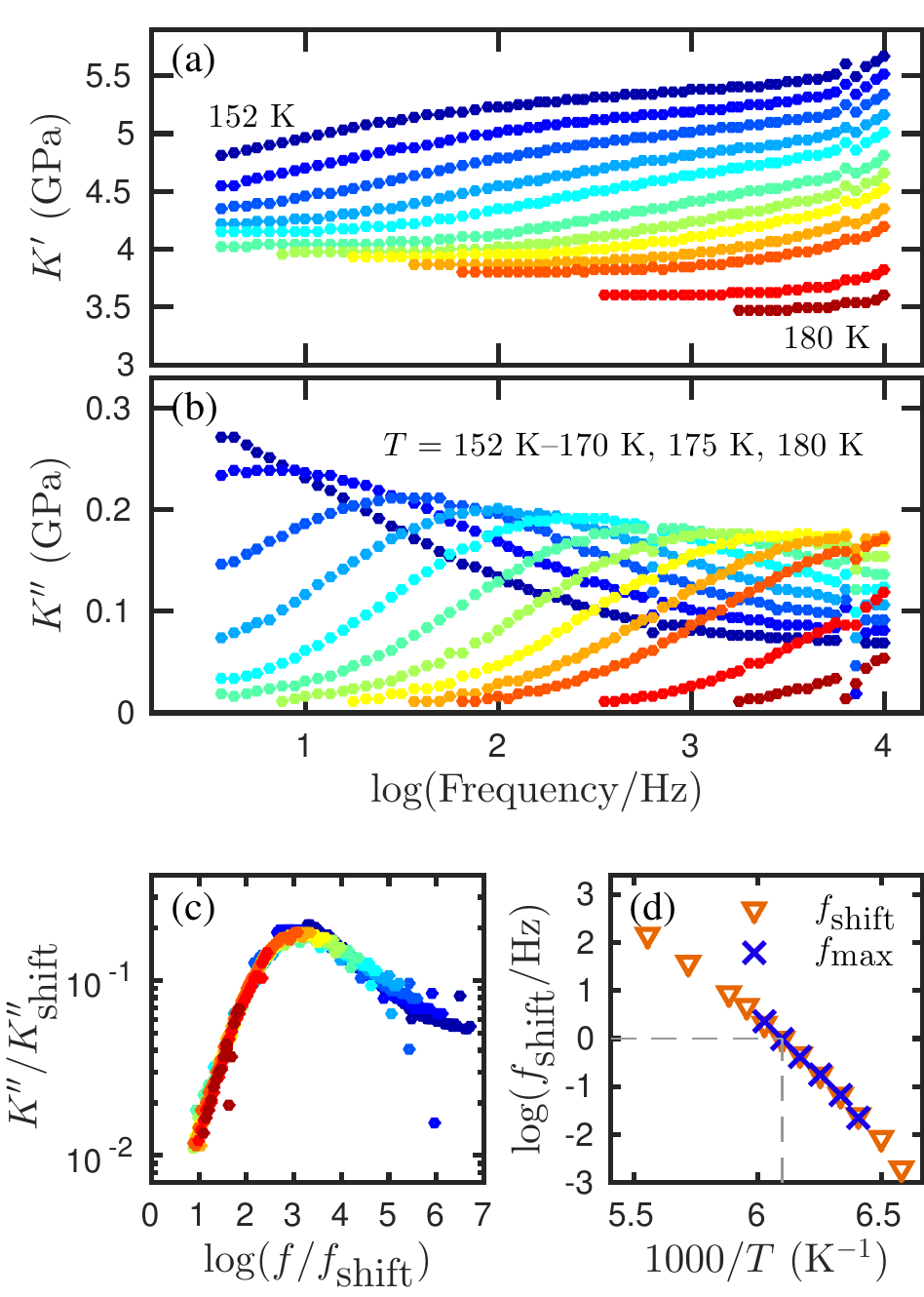} 
\caption{\label{2e1h_data}Complex bulk modulus of 2E1H at several
  temperatures. (a) Real part of the bulk modulus. (b) Imaginary part
  of the bulk modulus. (c) Master curve of the imaginary part of the
  bulk modulus data on a logarithmic scale. The data are shifted on
  both axes to match the data at 164~K. (d) The shift factors as a
  function of inverse temperature showing a non-Arrhenius evolution of
  the characteristic time scale of the bulk modulus. The reference
  (164~K) is marked by the dashed lines. For comparison, we show the
  peak frequencies for the temperatures where the peak is in the
  frequency window, shifted to overlap at the reference temperature. }
\end{figure}

The bulk modulus spectra are shown in Fig.\ \ref{2e1h_data}. The
frequency range of this measurement is from 2~Hz to 10~kHz. We
clearly see the low-frequency liquid-like plateau at the high
temperatures (red curves) and the high-frequency elastic plateau of
the lower temperatures (blue curves) in the real part (Fig.\
\ref{2e1h_data}(a)). The transition from low- to high-frequency
plateau gives a peak in the imaginary part (Fig.\ \ref{2e1h_data}(b))
as required by the Kramers-Kr\"onig relations. The peak is the alpha
relaxation peak, which moves down in frequency from around 10~kHz at
170~K to around 10~Hz at 154~K. At the lowest temperatures we see the
onset of another peak (the beta relaxation) at the highest
frequencies. The emergence of a secondary process is more evident in
the master curve constructed in Fig.\ \ref{2e1h_data}(c) by shifting
each of the curves on the frequency axes to give the best overlap. The
shift factors are shown in Fig.\ \ref{2e1h_data}(d) together with the
loss peak frequencies for the temperatures where the peak is inside
the frequency window. There is excellent agreement between the two
characteristic frequencies in the overlapping temperature region, thus
confirming that the shifts made are reasonable.

\begin{figure}
\begin{center}
\includegraphics[width=8cm]{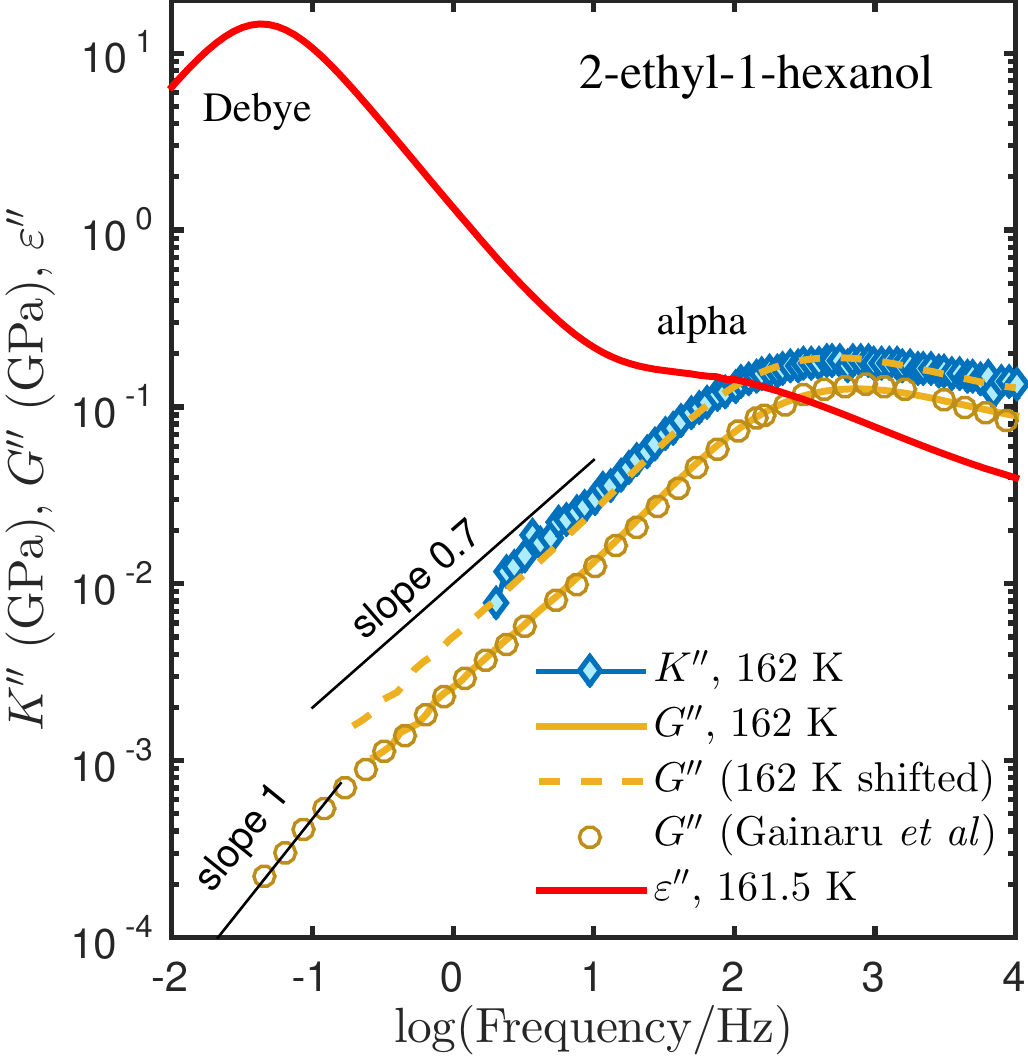} 
\end{center}
\caption{\label{2e1h_bsd}Comparison of the imaginary parts of the bulk
  modulus (blue diamonds), shear modulus (solid yellow line), and
  dielectric constant (solid red line). Shear modulus is compared to
  the master curve presented in Ref. \onlinecite{Gainaru2014} (yellow
  circles) and reproduces the earlier data. The dielectric data at a
  matching temperature are from Ref. \onlinecite{Jakobsen2008}. The
  signature of the short-chain polymer like behavior is a slope~$<1$
  on the low-frequency side of the shear modulus peak bending over to
  a purely viscous behavior (slope~$=1$) at $\sim0.1$~Hz, corresponding
  roughly to the loss-peak frequency of the Debye peak in the
  dielectric spectrum. The bulk modulus curve follows the
  low-frequency behavior of the shear modulus accurately, but the
  crossover to purely viscous behavior is below the resolution limit of
  the bulk modulus measurement.}
\end{figure}

Figure \ref{2e1h_bsd} shows the imaginary parts of bulk modulus $K$
(blue diamonds) and shear modulus $G$ (yellow line) at the same
temperature, 162~K. The shear modulus accurately reproduces the
previously published results \cite{Gainaru2014} shown in yellow
circles. For comparison, the dielectric curve at 161.5~K from Ref.\
\onlinecite{Jakobsen2008} is shown (red solid line). The dominant
feature in the dielectric spectrum is the Debye process characteristic
of mono-alcohols, where the alpha process is manifested as a
high-frequency shoulder placed roughly where the alpha peak in the
mechanical spectra appear. The recently established shear mechanical
signature of the Debye process \cite{Gainaru2014} shows up in the
shear spectra as a deviation from the pure viscous behavior observed
for non-associated liquids characterized by $G''(\omega) \propto
\omega$ (unity slope in a double-logarithmic plot). The crossover from
this intermediate low-frequency power-law behavior with $G''(\omega)
\propto \omega^{0.7}$ (slope 0.7 in a double-logarithmic plot) to the
terminal viscous flow with $G''(\omega) \propto \omega$ behavior is
located at a frequency close to the Debye peak frequency in the
dielectric spectrum.

The shifted shear mechanical curve (yellow dashed line) shows that the
bulk modulus displays the same deviation from pure viscous behavior as
the shear modulus. Although the resolution of the bulk modulus
measurement does not allow for showing a crossover to a pure viscous
behavior, the bulk modulus in non-associated liquids typically
displays a low-frequency purely viscous ($K'' \propto \omega$) behavior
\cite{Hecksher2013, Gundermann2014}; hence a terminal purely viscous
behavior is expected at frequencies and moduli below our current
resolution limit.

To appreciate our findings it is important to give a few experimental
details which is done in the following three paragraphs.

\begin{figure}[h!]
  \includegraphics[width=6.8cm]{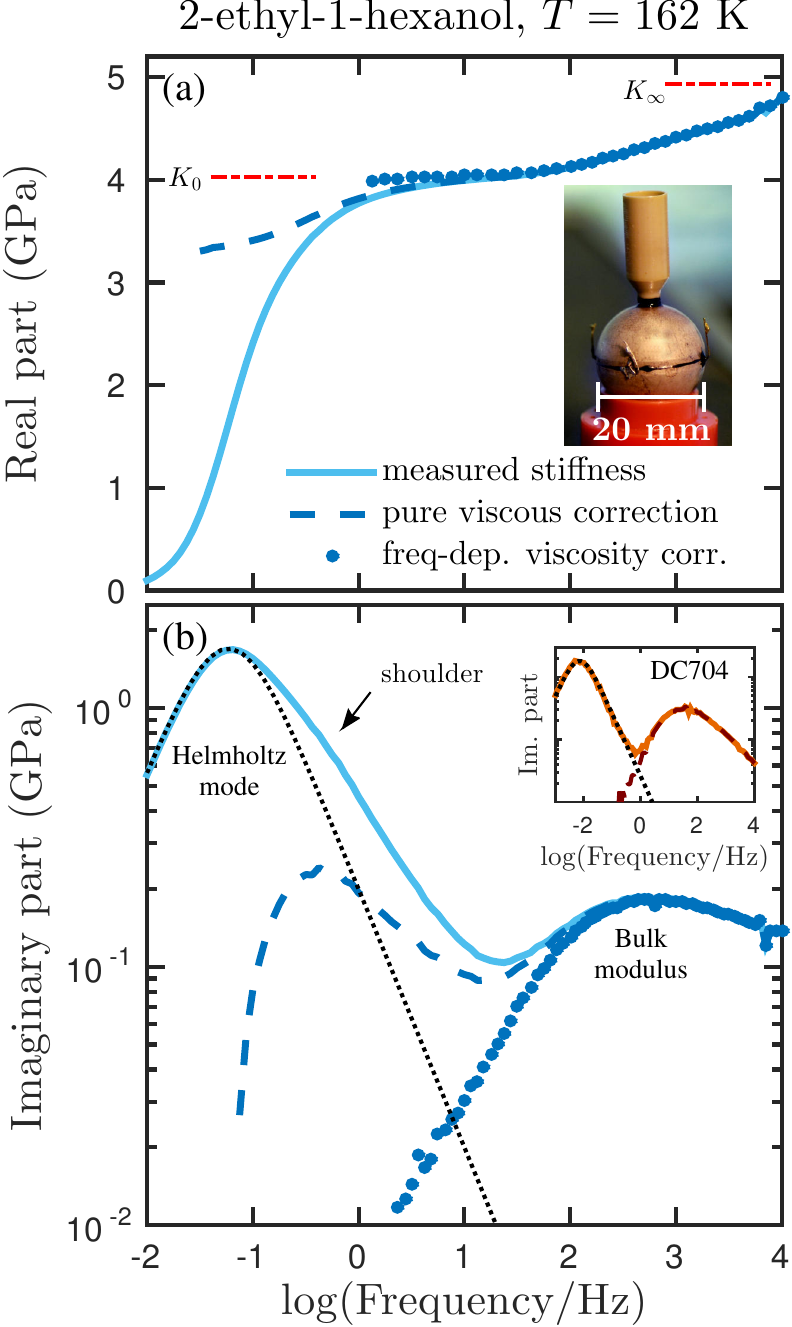} 
  \caption{\label{2e1h_cordat}Real (a) and imaginary (b) parts of the
    measured stiffness of 2E1H at 162~K. Inset of (a) is a picture of
    the measuring cell, the PBG \cite{Igarashi2008a,
      Christensen1994b,Hecksher2013}. At low frequencies, the real
    part of the measured stiffness goes to zero (light blue, solid
    line), because the liquid has time to flow in and out of the small
    filling tube. In the imaginary part, this is seen as a peak,
    denoted the Helmholtz mode. The simple, pure viscous correction is
    shown in both real and imaginary parts as the dashed blue line. It
    corresponds to ``subtracting'' the dotted black line in the
    imaginary part. While this procedure works perfectly in simple
    liquids, extending the frequency range of the bulk modulus by
    0.5-1 decade (see inset of (b), where this is shown for DC704), it
    leads to an extra peak in the imaginary part and a corresponding
    extra step in the real part in the present case. The apparent
    extra process comes from the ``shoulder'' indicated by an arrow in
    the figure. The shoulder is consistent with the slow polymer-like
    frequency dependence of the low-frequency side of the shear
    viscosity in 2E1H recently documented \cite{Gainaru2014}
    influencing the Poiseuille flow. Inserting the measured
    frequency-dependent shear modulus in the model (see Supplementary
    Materials) completely removes this shoulder and reveals the true
    bulk modulus (blue crosses).}
\end{figure}

The technique used for measuring the dynamics bulk modulus is based on
measuring the frequency-dependent capacitance of a piezo-electric
spherical shell of 20~mm diameter -- the Bulk Modulus Gauge (PBG)
\cite{Christensen1994b}. A picture of the PBG is shown as an inset of
Fig.\ \ref{2e1h_cordat}(a). When the PBG is empty it is free to deform
in response to an applied voltage, and we measure a ``free''
capacitance. When the PBG is filled with a liquid, the liquid
partially clamps the shell and, depending on the frequency, one
measures a lower capacitance. The dynamic bulk modulus can then be
inferred from the difference in capacitance between that of the freely
moving shell and the partially clamped shell. The translation from a
difference in capacitance to a bulk modulus of the liquid is somewhat
involved, because the signal is mixed with the mechanical properties
of the PBG itself and these effects needs to be
disentangled. Furthermore, the small filling hole in the piezo-ceramic
sphere leads to the liquid being able to flow \emph{out} of the PBG at
low frequencies. This is referred to as the ``Helmholtz mode''
\cite{Hecksher2013} which gives an extra feature in the measured
signal. Assuming that this flow can be described as a Poiseuille flow,
we can treat it as a hydraulic resistance depending only on the shear
viscosity of the liquid and compensate for the effect, making it
possible to extend the frequency window of the bulk modulus
measurement 0.5-1 decade. Details of this data extraction procedure is
given in the Supplementary Material and in
Refs. \onlinecite{Christensen1994b, Hecksher2010a, Hecksher2013}.

In Fig.\ \ref{2e1h_cordat} the procedure is illustrated. The real part
(Fig.\ \ref{2e1h_cordat}(a)) of the uncorrected stiffness goes to zero
at low frequencies, where the liquid flows in and out of the filling
pipe and thus does not resist the deformation of the PBG (light blue,
solid curve). This is the Helmholtz mode which in the imaginary part
(Fig.\ \ref{2e1h_cordat}(b)) appears as a prominent low frequency
peak. The assumption of a purely viscous flow leads to a functional
form (details are given in the Supplementary Material) of the peak
shown in the imaginary part as a black dotted line and labelled
``Helmholtz mode''. The peak in the 2E1H data clearly has a shoulder
in excess of this prediction, which when ``subtracted'' from the
measured stiffness leaves a small peak (dark blue, dashed line). The
inset of Fig.\ \ref{2e1h_cordat}(b) shows that the prediction for a
purely viscous flow holds and is easily ``subtracted'' from the
stiffness for a non-associated, molecular liquid (data from Ref.\
\onlinecite{Hecksher2013}).

The deviation from pure exponential behavior observed in the Helmholtz
mode here can, however, be accounted for by inserting the measured
\emph{frequency-dependent} shear viscosity (defined as $\eta =
G/(i\omega)$) instead of assuming a pure viscous flow. Then the
true bulk modulus (blue circles) is revealed; this correction lifts
the low-frequency real part to give a flat plateau and removes the
spurious extra peak in the imaginary part. This fact provides clear
and independent evidence that the low-frequency shear flow behavior of
2E1H differs both qualitatively and quantitatively from that of a
non-associated liquid.

Compared to poly-alcohols the relaxation strength of the bulk modulus
(and the shear modulus) of 2E1H is quite modest, roughly $1$~GPa at
the lowest temperatures. This relaxation strength is similar to what
has been found for the shear modulus of other mono-alcohols, for
instance $n$-propanol \cite{Kono1966}, 1-octanol and 2-octanol
\cite{Palombo2006}, and also what is typically found for the bulk and
shear modulus of non-associated, molecular liquids \cite{Hecksher2013,
  Gundermann2014}.  In glycerol \cite{Klieber2013} and propylene
glycol \cite{Maggi2008, Gundermann2014} the relaxation strength is a
factor of four higher.

We speculate that this fact may be understood in terms of the type of
molecular network formed in the poly-alcohols compared to the
mono-alcohols. Poly-alcohols can form branched hydrogen-bonded
networks which could lead to a stiffer structure and thus higher bulk
and shear moduli. The mono-alcohols, on the other hand, primarily form
linear structures (chains and/or rings). This presumably gives a
``looser'' structure, which is easier to deform and compress. The
signature of these structures in the shear spectra is a subtle
deviation from pure viscous behavior on the low-frequency side of the
alpha peak, similar to that of a short-chain polymer
\cite{Gainaru2014, Gainaru2014a, Hecksher2014, Adrjanowicz2015}. It is
not trivial, however, that the bulk modulus of mono-alcohols deviates
from that of non-associated liquids. The results presented here show
that the bulk modulus displays the same low frequency behavior as the
shear modulus, i.e., a non-trivial power law behavior at frequencies
below the alpha relaxation peak. Thus the dielectric Debye process
observed in mono-alcohols indeed does couple directly to the density
fluctuations in the liquid. In my opinion, these findings open up for
a less exotic explanation of the mysterious low-frequency Debye
process, possibly based on conventional liquid state theory via
density fluctuations.

\acknowledgements The author thanks Bo Jakobsen and Tage Christensen
for useful discussions and suggestions on the modeling and Jeppe Dyre
for input to and feedback on the manuscript. This work was sponsored
by the DNRF Grant no 61.

%

\pagebreak
\widetext
\begin{center}
\textbf{\large Supplementary Material: Linking the dielectric Debye process in 2-ethyl-1-hexanol to its density fluctuations.}
\end{center}
\setcounter{equation}{0}
\setcounter{figure}{0}
\setcounter{table}{0}
\setcounter{page}{1}
\makeatletter
\renewcommand{\theequation}{S\arabic{equation}}
\renewcommand{\thefigure}{S\arabic{figure}}
\renewcommand{\bibnumfmt}[1]{[S#1]}
\renewcommand{\citenumfont}[1]{S#1}
\section{Experimental details}

The sample, 2-ethyl-1-hexanol (2E1H), was purchased from Sigma-Aldrich
(purity $\geq 99.6$~\%) and used as received.

All bulk and shear modulus measurements were carried out in the same
experimental set-up, at the same temperatures (however, the shear
modulus measurement included some lower temperatures than the bulk
modulus measurement) and frequencies.

The set-up includes a custom-built closed-cycle cryostat able to keep
the temperature stable within about 1~mK. The temperature calibration
procedure benchmarks absolute temperature stability to better than
0.1~K, however the actual absolute temperature stability may be
slightly worse, because it depends on temperature gradients in the
cryostat, i.e., the thermal contact between sample cell and the stick,
and the thermal contact between the stick and the inner walls of the
cryostat chamber. This can vary between different types of cells (due
to slightly different positions in the cryostat, difference in masses
and/or materials) and from measurement to measurement (due to
different handling).

The electronics of the set-up consist of a custom-built generator for
frequencies up to 100~Hz, a HP 3458A multimeter (measuring at low
frequencies) and an Agilent E4980A Precision LCR meter (measuring
frequencies up to 2~MHz).

Further details of the experimental set-up is given in
Refs. \onlinecite{S_Igarashi2008a, S_Igarashi2008b}. Details of measuring
techniques in given Ref. \onlinecite{S_Christensen1995} (shear modulus
technique) and Refs. \onlinecite{S_christensen1994b, S_Hecksher2013} (bulk
modulus technique).

\section{Shear mechanical data}

Shear modulus data for 2E1H was published in
Refs. \onlinecite{S_Jakobsen2008, S_Gainaru2014}, but rather than using
the old data we decided to measure again. This was done for several
reasons: 1) previous measurements were carried out at slightly
different temperatures 2) measurements were performed in a different
experimental set-up 3) it was a ``cleaner'' implementation when data
points are at exactly the same frequencies (as in the bulk modulus
measurement) instead of shifting a master curve in frequency and then
interpolating between frequencies to match the bulk modulus
measurement.

\begin{figure}[h!]
\includegraphics[width=7.9cm]{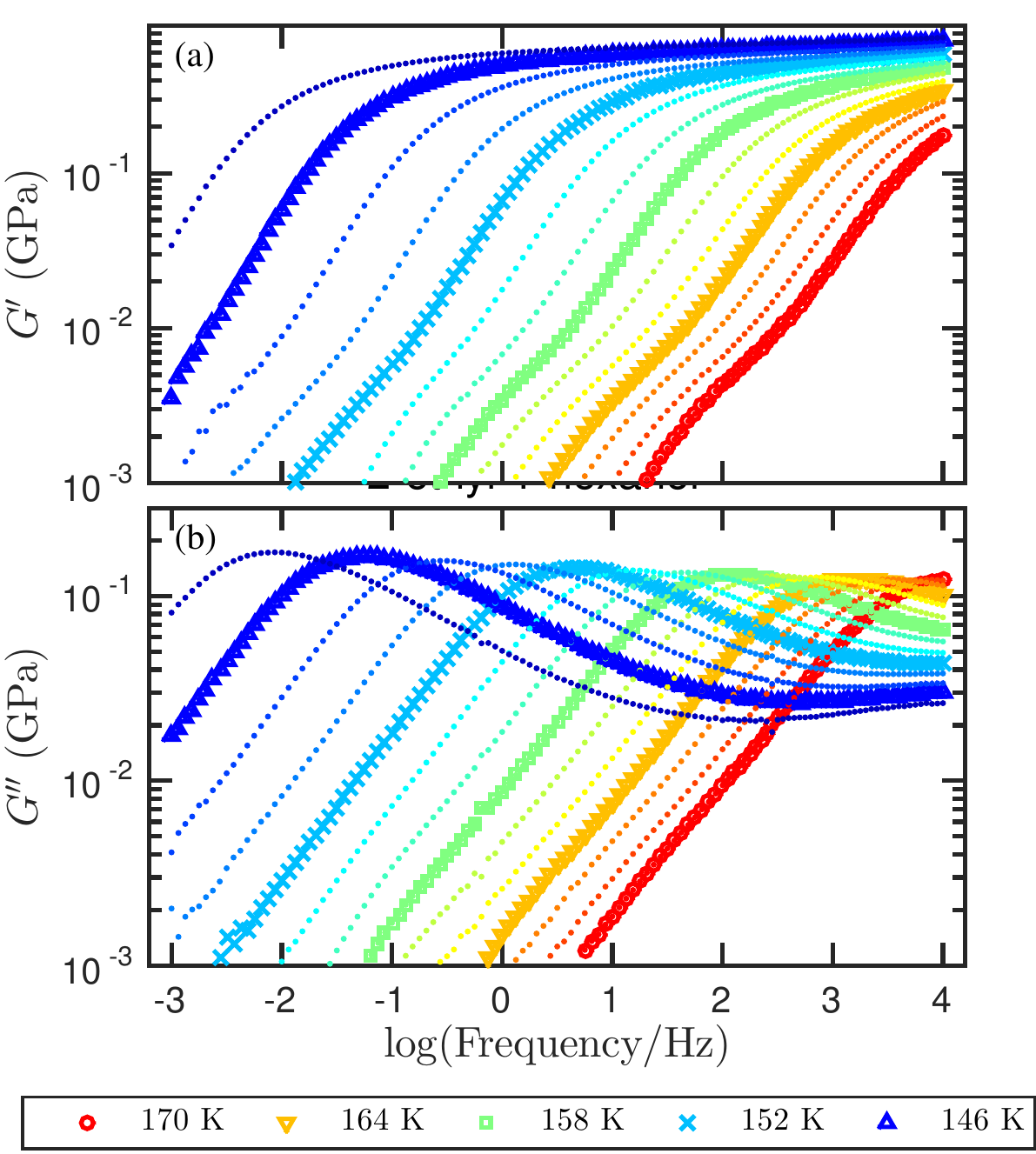} 
\caption{\label{shear2e1h}Real (a) and imaginary (b) part of the shear
  modulus of 2E1H at temperatures from 170~K to 144~K in
  steps of 2~K. }
\end{figure}

Shear modulus was thus re-measured in the same experimental set-up as
the bulk modulus, changing only the measuring cell while the cryostat,
cryostat stick, and electronics were identical under the two
measurements. Data are shown in Fig.\ \ref{shear2e1h}. The data agree
well with previously published shear modulus data \cite{S_Jakobsen2008,
  S_Gainaru2014} for 2E1H.

\section{Model for the PBG}

This section gives the background for the data treatment of the bulk
modulus data and is essential to the conclusions drawn. 

The PBG can be modeled by an electrical diagram, where each element
represents a particular property of the PBG \cite{S_Hecksher2013}. Such
a diagram is essentially a simple way of constructing the constitutive
equations of the PBG in a physically consistent manner.

\begin{figure}
\includegraphics[width=8cm]{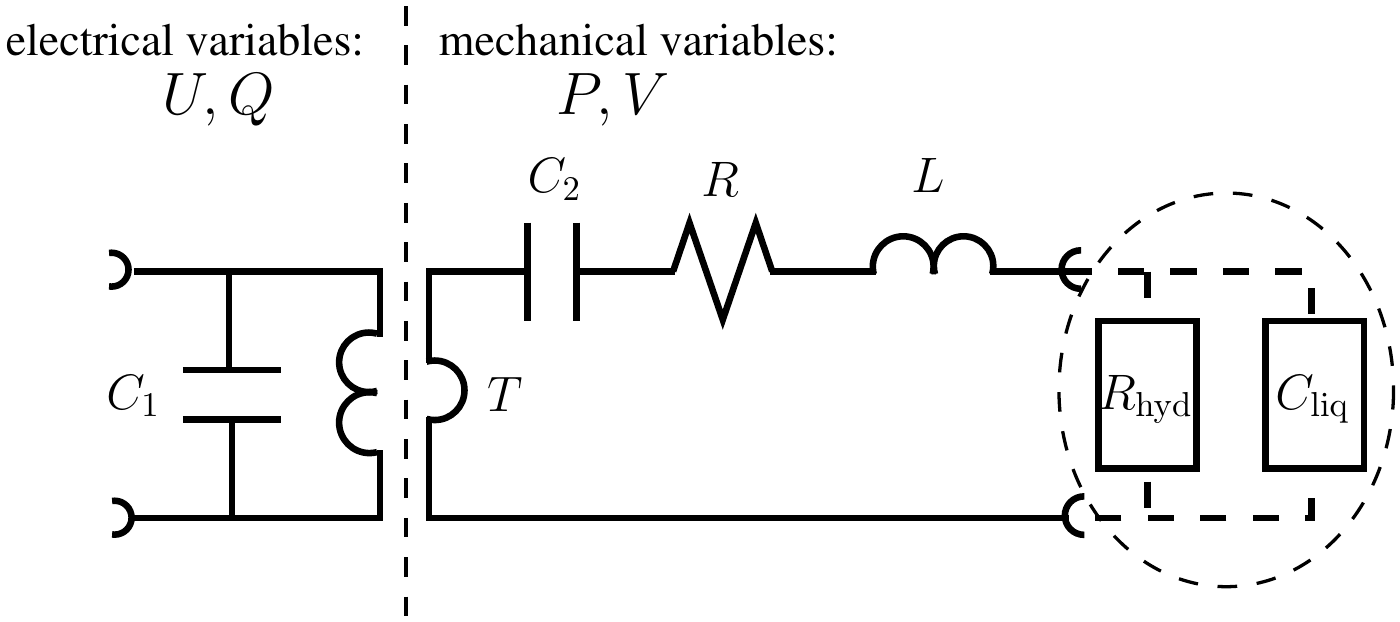}
\caption{\label{PBGmodel}Model of the PBG. The transformer (also
  marked by a vertical dashed line) separates the electrical and the
  mechanical side of the model. The encircled area shows the part that
  is only present when the PBG is filled. An empty transducer
  corresponds to the mechanical port of the model being shorted. The
  electrical side (left of the transformer, $T$) the capacitance $C_1$
  models the electrical capacitance of the piezo-electric ceramic
  shell. On the mechanical side (right of the transformer) the
  $RCL$-circuit models the mechanical properties of the ceramics. The
  transformer models the piezo-electric conversion of the applied
  electrical field to mechanical displacement. When there is liquid in
  the PBG, there is an extra (complex) capacitance due to the liquid
  ($C_\textrm{liq}$), and an extra (complex) resistor
  ($R_\textrm{hyd}$) modeling the flow in the filling pipe. }
\end{figure}

Figure \ref{PBGmodel} shows the electrical diagram model of the
PBG. The model has an electrical side (left) and a mechanical side
(right) where the volume $V$ plays the role of electrical charge $Q$
and the pressure $P$ is the equivalent of the voltage $U$ on the
electrical side.  The transformer (separating the two sides of the
diagram) represents the piezo-electric conversion of electrical
voltage to a mechanical displacement. The capacitance on the
electrical side $C_1$ models the electrical capacitance of the
piezo-electric ceramic shell. On the mechanical side, the
$RCL$-circuit models the mechanical properties of the ceramics. When
there is liquid in the transducer, an extra (complex) capacitance due
to the liquid ($C_\textrm{liq}$), and an extra (complex) resistor
($R_\textrm{hyd}$) modeling the flow in the filling pipe is added to
the model. 

Using the simple rules for adding electrical network elements
(impedances added in series, admittances added in parallel), we arrive
at the following expression for the measured capacitance, i.e., the
response when measured at the electrical side of the model
\begin{equation}\label{pbgmodres}
  C_\textrm{m}(\omega) = C_1 + T^2 \frac{1}{\frac{1}{C_2}+i\omega R -
    \omega^2 L + \left[ \frac{1}{\frac{1}{i\omega R_h} + C_\textrm{liq}}
    \right]} \,,
\end{equation}
where the expression in the square brackets only is present if the PBG
is filled with liquid.

Equation (\ref{pbgmodres}) can be rewritten in some convenient
variables: the clamped (high-frequency limit) capacitance
$C_{cl}=C_1$, the free (low-frequency limit) capacitance
$C_{fr}=C_1+T^2C_2$, the resonance frequency $\omega_0 =
\sqrt{1/LC_2}$, and the quality factor $Q=1/R \sqrt{L/C_2}$. The
rewritten expression is then
\begin{equation}\label{pbgmod2}
  C_\textrm{m}(\omega)  = C_{cl} + \frac{C_{fr} - C_{cl}}{1 +
    i\frac{\omega}{\omega_0} \frac{1}{Q} - \frac{\omega^2}{\omega_0^2}
    + \left[ \frac{C_2}{\frac{1}{i\omega R_\textrm{hyd}} +
        C_\textrm{liq}} \right]} \,. 
\end{equation}

\begin{figure*}
  \includegraphics[width=16cm]{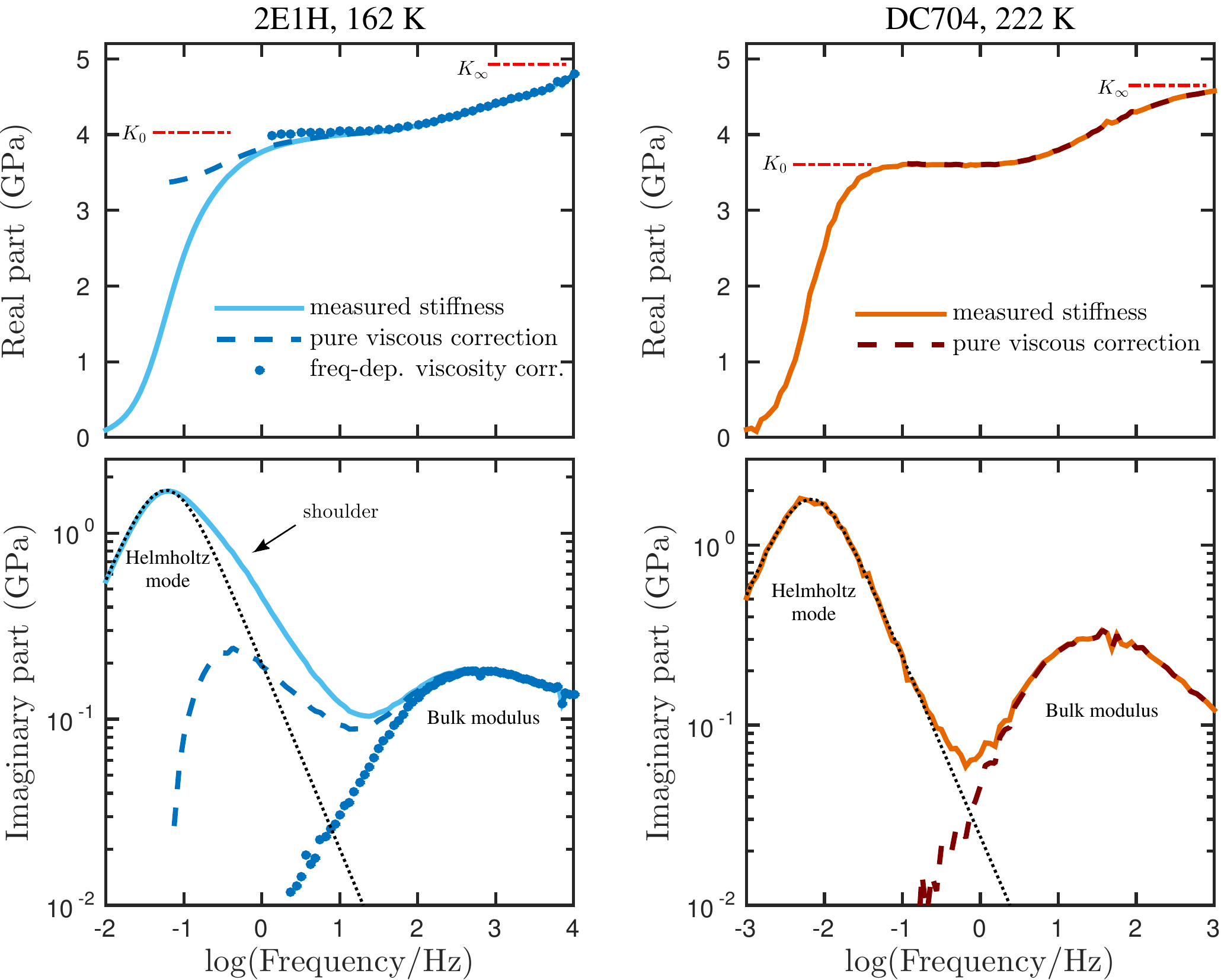} 
  \caption{\label{2e1h_cordatSM}Real and imaginary parts of the measured
    stiffness of 2-ethyl-1-hexanol (2E1H) and the silicone oil,
    tetraphenyl-tetramethyl-trisiloxane (DC704). At low frequencies,
    the real part of the measured stiffness goes to zero (solid line),
    because the liquid has time to flow in and out of the small
    filling tube and thus there is no resistance to the deformation of
    the piezo-ceramic shell. In the imaginary part this is seen as a
    peak, the Helmholtz mode. In the real part the limiting low- and
    high-frequency moduli, $K_0$ and $K_\infty$ are indicated by red
    dashed lines. The simple, purely viscous correction (see
    Eq. (\ref{pbgmodres})) is shown in both real and imaginary parts
    as the dashed line. It corresponds to the ``subtraction'' of the
    dotted black line in the imaginary part. While this procedure
    works perfectly in non-associated molecular liquids, extending the
    frequency range of the bulk modulus by approximately 1 decade
    (shown here for DC704), it clearly leads to an extra peak in the
    imaginary part and a corresponding extra step in the real part in
    the case of 2E1H. This apparent extra process comes from the
    ``shoulder'' indicated by an arrow in the imaginary part of the
    2E1H spectrum. The shoulder is consistent with the slow
    polymer-like frequency dependence of the shear viscosity in 2E1H
    recently documented \cite{S_Gainaru2014} influencing the Poiseuille
    flow. Inserting the measured frequency-dependent shear modulus in
    the model (Eq. (\ref{bmcorr})) completely removes this shoulder
    and reveals the true bulk modulus (blue crosses).}
\end{figure*}

The hydrodynamic flow resistance $R_\textrm{hyd}$ is proportional to
the shear viscosity, $\eta_G$. Assuming a Poiseuille flow (see
Sec. \ref{poiseuille} below), the factor of proportionality is given
by
\begin{equation}\label{geom}
  A = \frac{8L}{\pi a^4}\,,
\end{equation}
where $L$ is the length of the pipe and $a$ is the radius. Inserting
this and rearranging Eq.\ (\ref{pbgmod2}) to isolate $C_\textrm{liq}$
(which is the signal we are after), we arrive at
\begin{equation}
  \begin{split}
    C_\textrm{liq}(\omega) & = C_2\left(
      F^{-1}-1-i\frac{\omega}{\omega_0}\frac{1}{Q} +
      \frac{\omega^2}{\omega_0^2} \right)^{-1} - \frac{1}{i\omega A
      \eta_G} \\
  \end{split}
\end{equation}
where $F=\frac{C_\textrm{m}(\omega)-C_{cl}}{C_{fr}-C_{cl}}$. 

For liquids that display a ``simple'' low-frequency behavior in the
shear modulus (e.g., most non-associated molecular liquids), it is
sufficient to plug in the DC viscosity in the model (corresponding to
a pure resistor in the network in place of the $R_\textrm{hyd}$-box in
Fig.\ \ref{PBGmodel}). A more sophisticated model takes the
frequency-dependence of the viscosity into account. This could either
be done by putting in a more complicated model for $R_\textrm{hyd}$,
but one could also plug in the actual measured shear viscosity. Since
$\eta_G = \frac{G}{i\omega}$, where $G$ is the complex shear modulus
we finally arrive at
\begin{equation}\label{bmcorr}
  C_\textrm{liq}(\omega) = C_2\left( F^{-1} - 1 -
    i\frac{\omega}{\omega_0} \frac{1}{Q} +
    \frac{\omega^2}{\omega_0^2} \right)^{-1} - \frac{1}{AG} \,. 
\end{equation}
The bulk modulus (with the ``subtracted'' hydrodynamic flow in the
filling pipe) is obtained as follows
\begin{equation}
  K_S(\omega) = \frac{V}{C_\textrm{liq}(\omega)}\,,
\end{equation}
where $C_\textrm{liq}$ is given by Eq.\ (\ref{bmcorr}).

The procedure is illustrated in Fig.\ \ref{2e1h_cordatSM}, where both
the pure viscous correction and the correction including the
frequency-dependence of the viscosity is shown for 2E1H. For
comparison, we show how the pure viscous correction works for a
non-associated molecular liquid, the commercial silicone oil DC704
\cite{S_Hecksher2013}.

\section{Comments on the Poiseuille flow assumption}\label{poiseuille}

In fluid dynamics, the Poiseuille law relates the flowrate, $\dot{V}$
to the pressure drop $\delta P$ over the pipe
\begin{equation}\label{eqP}
  \dot{V} = R_\textrm{hyd}\delta P\,,
\end{equation}
where the hydrodynamic resistance $R_\textrm{hyd}$ inversely
proportional to the fluid's viscosity, $\eta$. The constant of
proportionality is the geometrical constant given by Eq.\
(\ref{geom}).

The assumptions of Eq.\ (\ref{eqP}) are that 1) the liquid is
incompressible and Newtonian, 2) the flow is laminar, 3) there is no
acceleration of fluid in the pipe and 4) the pipe has constant
circular cross-section and the length of the pipe is substantially
longer than its radius.

In our case we do not have a constant flow, but a pulsating flow where
the frequency of the pulsations varies from 1~mHz to 10~kHz. Of course
assumption of zero acceleration no longer holds, but the requirement
of a laminar flow translates into the frequency of pulsations being
sufficiently low so that a parabolic velocity profile has time to
develop during each cycle. In that case the Poiseuille equation hold
to a good approximation. This is fulfilled when Womersley number,
$\alpha$, is small. Wormersley number is given by
\begin{equation}
  \alpha = a \sqrt{\frac{\omega\rho}{\eta(\omega)}}\,.
\end{equation}
where $\omega$ is the angular frequency, $\rho$ is the density, $\eta$
is the frequency-dependent viscosity and $a$ is the radius of the
pipe.

\begin{figure}
\includegraphics[width=8cm]{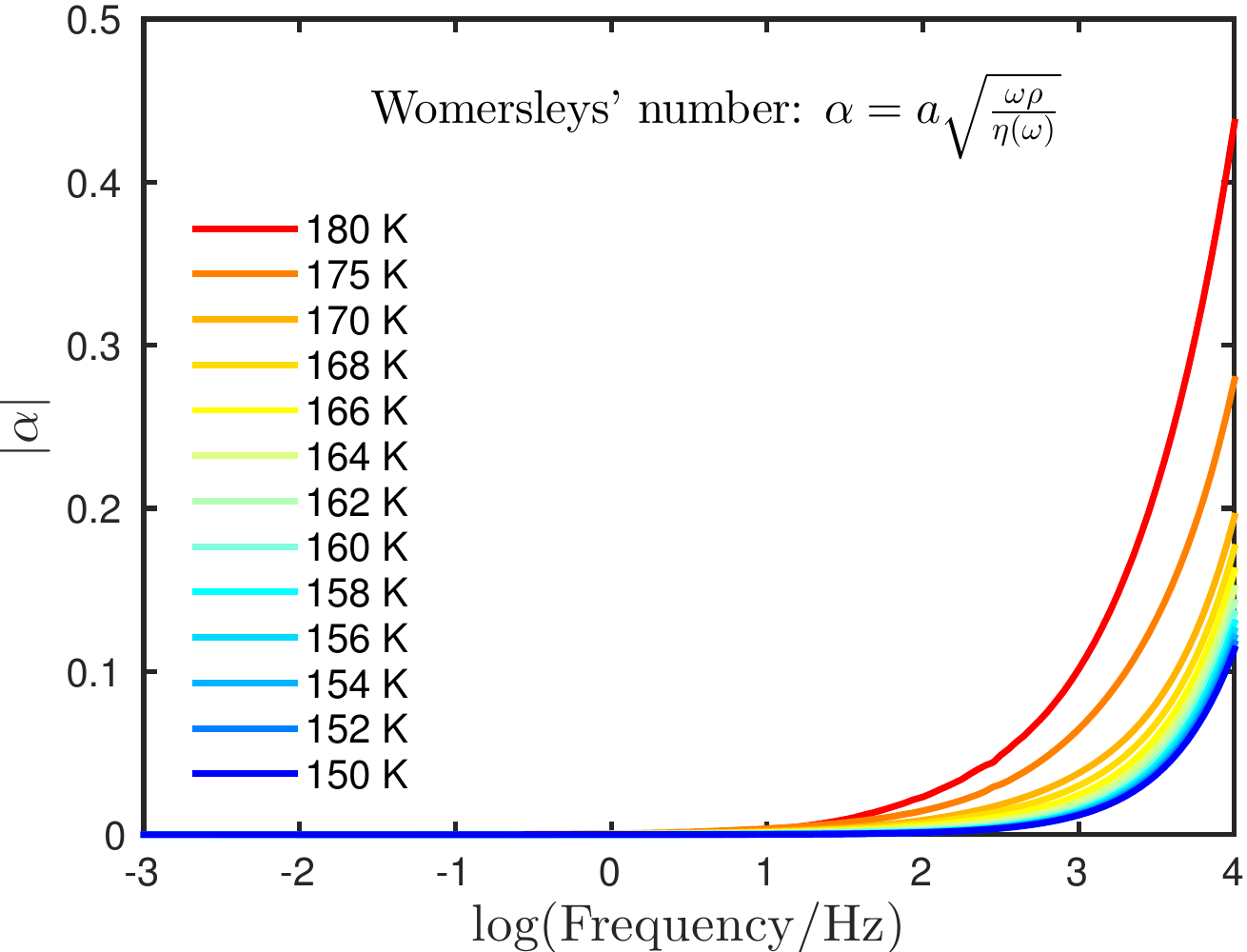}
\caption{\label{womersley}Womersley number calculated as a function of
  frequency for 2E1H at several temperatures. When $\alpha$ is small
  ($<1$), it means the frequency of pulsations is sufficiently low
  that a parabolic velocity profile has time to develop during each
  cycle, and the flow will be very nearly in phase with the pressure
  gradient. The flow will then be given by Poiseuille's law to a good
  approximation, using the instantaneous pressure gradient.}
\end{figure}

In Fig.\ \ref{womersley} the Womersley number is shown as a function
of frequency and temperature in the case of 2E1H and clearly shows
that $|\alpha|<1$ at all temperature in the relevant frequency range,
and thus assumptions two and three are met.

The filling pipe is cylindrical, the radius is approximately
$a\approx1.5$~mm, while the length is approximately
$L\approx4$~mm. The requirement that $a/L\ll1$ is maybe not completely
met. However, when in a pulsating flow the problem is smaller at high
frequencies. In agreement with this, we observe that the correction
building on the Poiseuille flow assumption gradualy breaks down for
low frequencies (i.e., for frequencies lower than the peak frequency
of the Helmholtz mode, see Fig.\ \ref{2e1h_cordatSM}).

The Poiseuille equation has been shown to work (surprisingly) well for
supercooled molecular liquids at the frequency ranges and pipe
dimensions explored here \cite{S_Hecksher2013}.

\section{Determining the geometric factor}

The geometric factor in Eq.\ (\ref{geom}) that enters the calculation
of the Poiseuille flow correction consist of the length, $L$, and
diameter, $a$, of the filling tube. In principle, these quantities can
be determined by measuring directly the dimension of the tube. In
practice, this is not so easy since the filling tube is hidden at the
bottom of a larger liquid reservoir, and so it is difficult measure
the tiny dimensions accurately inside measuring cell.

\begin{figure}
\includegraphics[width=6cm]{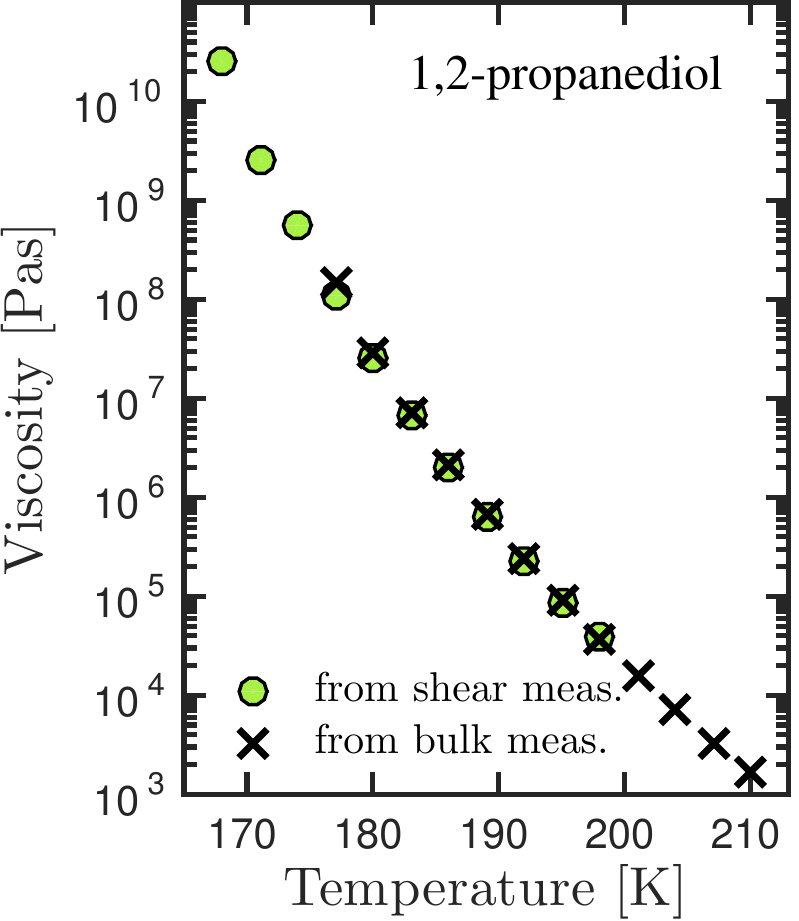} 
\caption{\label{shearviscPD}Shearviscosity of 1,2-propanediol measured
  directly as the low-frequency plateau of the real part of the
  complex shearviscoity, $\eta=G/i\omega$, and as inferred from the
  Poiseuille flow in the bulk transducer described in the text (for
  more details, see Ref.\ \cite{S_Hecksher2013}). Data are from
  Ref. \cite{S_Gundermann2014}. This determines the geometric factor of
  the translation between hydraulic resistance and shear viscosity for
  that particular transducer (named q10 for internal reference) to $A
  = 3.6 \times 10^{9}$~m$^{-3}$.}
\end{figure}

Instead we `calibrate' the geometrical constant with another set of
shear and bulk modulus data measured in the same experimental set up,
and -- in the case of the bulk modulus -- in the same measuring
cell. In Fig.\ \ref{shearviscPD} we show the static shear viscosity as
determined from the low-frequency plateau value of the
frequency-dependent viscosity measured in the PSG (green circles) as
well as the static shear viscosity determined by the Helmholtz mode
\cite{S_Hecksher2013} with a geometric factor adjusted, so the two
curves coincide. The geometric factor used for correcting the
low-frequency side of the bulk modulus spectra was $A = 3.6 \times
10^{9}$~m$^{-3}$. This values agrees qualitatively with the number
found when inserting the values for radius and length stated in Sec.\
\ref{poiseuille} above in Eq.\ (\ref{geom}).

\end{document}